\documentclass[useAMS,usenatbib]{mn2e}

\usepackage{psfig, epsf, epsfig}

\title[Forming galactic nuclei]{A model for 
GN-z11: top-heavy stellar initial mass functions in 
forming galactic nuclei and ultra-compact dwarfs}
\author[K. Bekki and  T. Tsujimoto]
{Kenji Bekki${}^1$\thanks{E-mail:
kenji.bekki@uwa.edu.au} 
and
Takuji Tsujimoto${}^2$  \\
${}^1$ICRAR M468
The University of Western Australia
35 Stirling Hwy, Crawley
Western Australia 6009, Australia  \\
${}^2$ National Astronomical Observatory of Japan, Mitaka-shi, Tokyo 181-8588, Japan
}

\begin{document}

\date{Accepted, Received 2005 February 20; in original form }

\pagerange{\pageref{firstpage}--\pageref{lastpage}} \pubyear{2005}

\maketitle

\label{firstpage}

\begin{abstract}

Recent JWST observations of the z=10.6 galaxy  GN-z11 have revealed
a  very high
gas-phase nitrogen abundance (higher than four times the solar value),
a very small half-light radius 
($\approx 60$ pc), and 
a large stellar mass ($M_{\rm s} \approx 10^9 {\rm M}_{\odot}$) for its size.
We consider that this object is a forming galactic nucleus or ultra-compact dwarf galaxy
rather than a proto globular cluster,
and thereby investigate the chemical abundance pattern using one-zone chemical evolution
models.
The principal results of the models are as follows. The observed $\log {\rm (N/O)} > -0.24$,
$\log {\rm (C/O)}>-0.78$, and 12+$\log {\rm (O/H)} \approx 7.8$ can be self-consistently
reproduced by the models both with very short star formation timescales ($< 10^7$ yr)
and with top-heavy stellar initial mass functions (IMFs).
The adopted assumption of no chemical enrichment by 
massive ($m>25 {\rm M}_{\odot}$) core collapse supernovae  (CCSNe)
is also important for the reproduction of
high gas-phase  $\log {\rm (N/O)}$,
because such CCSNe
can decrease high $\log {\rm (N/O)}$ of gas polluted by OB and Wolf-Rayet stars.
GN-z11 can have a significant fraction ($>0.5$) of nitrogen-rich ([N/Fe]$>0.5$) stars,
which implies a possible link between nitrogen-rich stellar populations 
of the inner Galaxy and  giant elliptical galaxies and high-$z$ objects with high gas-phase
$\log {\rm (N/O)}$  like GN-z11.
\end{abstract}

\begin{keywords}
galaxies:evolution --
infrared:galaxies  --
stars:formation  
\end{keywords}

\section{Introduction}

A significant fraction of dwarf and late-type spiral galaxies in various 
environments  are observed to contain
stellar galactic nuclei or ``nuclear star clusters''
(e.g., Sandage \& Binggeli 1984; B\"oker et al. 2002; C\^ote et al. 2006).
The physical origin of the stellar nuclei and their evolutionary
links to massive black holes (MBHs) dominating the central regions of massive early-type
galaxies are yet to be fully understood (e.g.,  Graham \& Spitler 2008; Antonini et al. 2015). 
The observed very small half-light radii ($R_{\rm h}<100$ pc) and large stellar masses
($10^6 \le M_{\rm s}/{\rm M}_{\odot} \le 10^8$) in ultra-compact dwarf (UCD)
galaxies (e.g., Drinkwater et al. 2003; Mieske et al. 2008) have been extensively
discussed in the context of their transformation  from massive  nucleated dwarf galaxies
(e.g., Bekki et al. 2001;  Pfeffer et al. 2014). 
Recent spectroscopic studies of a few UCDs (e.g., M60-UCD1 and UCD3)
have confirmed the possible presence of MBHs
(Seth et al. 2014; Afanasiev et al. 2018), 
which implies that there is a link between UCD and MBH
formation. 
The nitrogen abundance  of M60-UCD1 is observed to be strongly
enhanced with  [N/Fe]$=0.61\pm 0.04$ (Strader et al. 2013).

Recent JWST observations of the $z=10.6$ galaxy GN-z11 first identified by Bouwens et al. (2010)
have revealed (i) an intriguing morphology with 
a central point source and an  outer diffuse stellar envelope,
(ii) very compact $R_{\rm h}$ ($=64 \pm 20$ pc),
(iii) large stellar mass ($\log (M_{\rm s}/M_{\odot})=9.1^{+ 0.3}_{-0.4}$) for its size,
and (iv) unusually large gas-phase nitrogen abundance
with $\log {\rm (N/O)} > -0.24$  (e.g., Bunker et al., 2023;
Tacchella et al. 2023; Cameron et al. 2023).
A number of authors have already discussed the origin of the high nitrogen abundance
in the context of ejection of nitrogen-rich  gas from tidal destruction of 
massive stars in a star cluster (Cameron et al. 2023), the possible presence of
supermassive stars (Charbonnel et al. 2023),  chemical enrichment by Wolf-Rayet (WR) stars
(Watanabe et al. 2023), and globular clusters (GCs) at their birth (Senchyna et al. 2023; S23).
Given that the observed morphology and structure of GN-z11 are reminiscent of massive nucleated dwarf
galaxies or UCDs,  it is quite reasonable to discuss the high $\log {\rm (N/O)}$ 
in the context of stellar nucleus/UCD  formation at high $z$.

\begin{table}
\centering
\begin{minipage}{85mm}
\caption{The values of model parameters.}
\begin{tabular}{lllll}
Model ID  & $\alpha$ & $t_{\rm sf}$ (yr) & $t_{\rm inf}$ (Myr) & $f_{\rm r}$    \\
M1  & 1.55 & $6.6 \times 10^6$  & 1 & 0.03  \\
M2  & 1.55 &  $6.7 \times 10^7$ & 1 & 0.03   \\
M3  & 1.15 &   $6.7 \times 10^6$  & 1 & 0.03  \\
M4  & 1.95 &  $6.6 \times 10^6$  & 1 & 0.03 \\
M5  & 2.35 &  $6.6 \times 10^6$ & 1 & 0.03  \\
M6  & 2.75 &  $6.6 \times 10^6$ & 1 & 0.03  \\
M7  & 1.55 &  $6.6 \times 10^6$ & 1 & 0.1  \\
M8  & 1.55 &  $6.6 \times 10^6$ & 1 & 0.2  \\
M9  & 1.55 &  $6.6 \times 10^6$ & 1 & 1.0  \\
M10  & 2.35 &   $6.6 \times 10^6$  & 300 & 1.0 \\
\end{tabular}
\end{minipage}
\end{table}

\begin{figure*}
\psfig{file=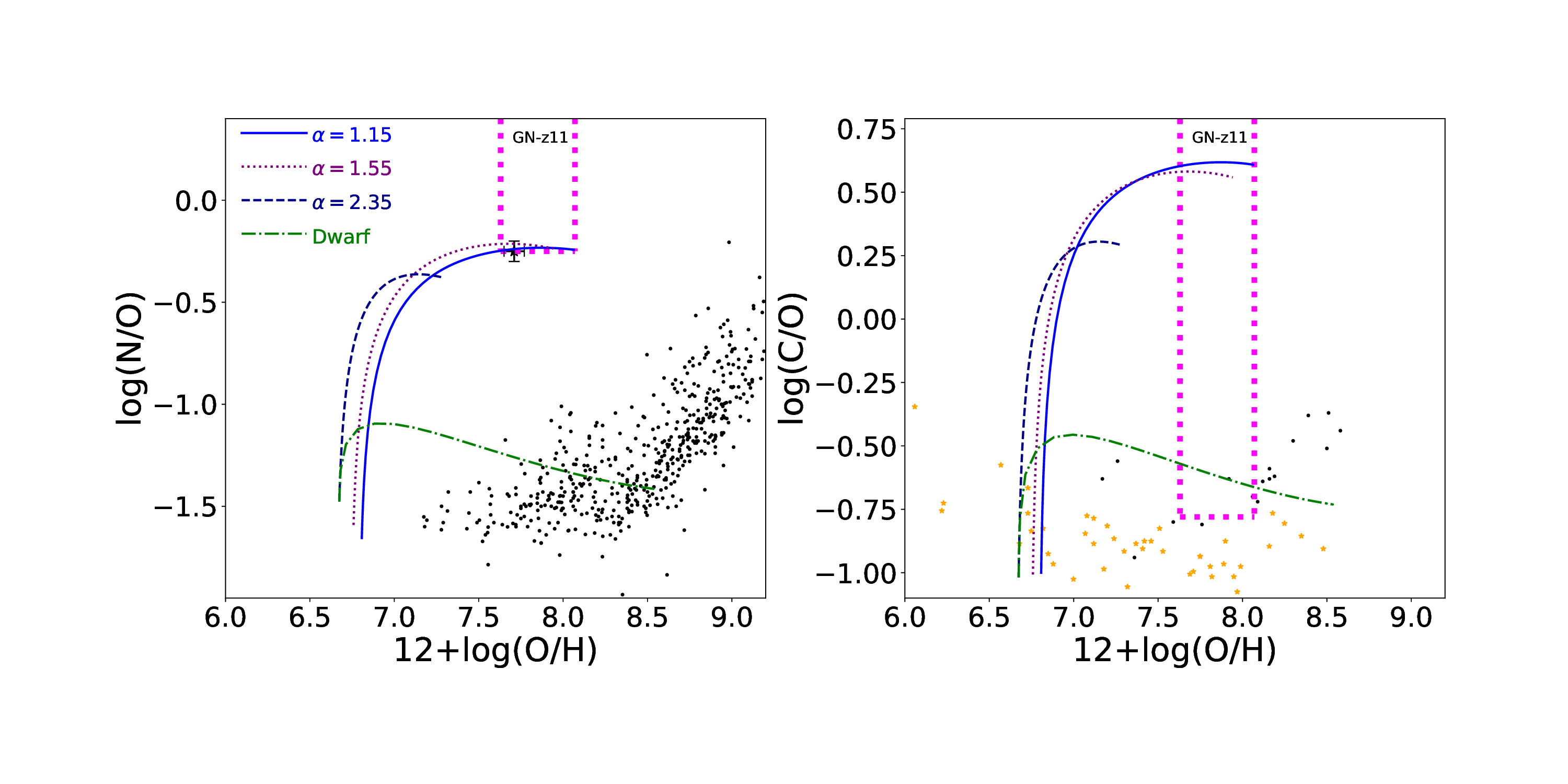,width=19.0cm}
\caption{
Time evolution of the model M3  with $\alpha=1.15$ (blue solid),
M1 with $\alpha=1.55$ (purple dotted), 
M$\alpha=2.35$ (Salpeter, dark-blue dashed), 
and M10  (green dot-dashed, referred to as the ``dwarf'' model) 
on the $\log {\rm 12 + (O/H)}-\log {\rm (N/O)}$ (left)
and $\log {\rm 12 + (O/H)}-\log {\rm (C/O)}$ planes (right).
For comparison, the longer term ($10^8$ yr) evolution is
shown for the dwarf model (M10).
The observational data for the low-metallicity extragalactic  HII regions from Pettini et al. (2008)
are plotted by small black dots in the two panels. 
The observed $\log {\rm (C/O)}$ of the Galactic halo stars from
Fabbian et al. (2009) are plotted by small orange stars in the right panel only.
The observed (fiducial) limits of these abundances 
(Cameron et al. 2023) are indicated by magenta dotted lines.
The  model for $\log {\rm (N/O)}$  by S23.
is shown by a black star with error bars
}
\label{Figure. 1}
\end{figure*}

Je{\v{r}}{\'a}bkov{\'a} et al. (2017) pointed out that forming
$z=9$ UCDs with $M_{\rm s}=10^8 {\rm M}_{\odot}$
and a top-heavy
stellar initial mass function (IMF) with the slope ($\alpha$) as flat as 1.1
(2.35 for the Salpeter type)
can be  detected by the JWST NIRcam (e.g., F115W)
with $S/N > 7$ for a $\approx 3$ hour exposure time.
GC formation  models by Bekki \& Chiba (2007)
show that if the IMF is top-heavy with $\alpha \approx 1.5$,
gas-phase nitrogen abundance 
can be rather high ([N/Fe]$>0.5$) for a large mass
fraction of the intra-cluster gas due to pollution by massive stars.
Although the models by Bekki \& Chiba (2007)
would be relevant to a star cluster with
$M_{\rm s} \approx 10^7 {\rm M}_{\odot}$ and $\log {\rm (N/O)} \approx -0.21$
found in the Sunburst Arc (Pascale et al. 2023),
they need to be revised to discuss the origin of GN-Z11
that is much  more massive than GCs.

The purpose of this paper is to show that 
the observed high gas-phase $\log {\rm (N/O})$ of GN-z11
can be consistent with the formation of massive stellar nuclei
or UCDs  with top-heavy IMFs.
Using one-zone chemical evolution models for GN-z11, we investigate in what physical conditions
the observed  high log(N/O) and log(C/O) can be reproduced  self-consistently
by our models.
We particularly investigate how (i) the IMF, (ii) gas consumption/infall timescales,
and (iii) mixing process of stellar winds and CCSNe
determine the  chemical abundance
patters of gas and stars in our models.

\section{The model}

We investigate the chemical abundances of gas and stars in GN-z11 using one-zone
chemical evolution models adopted in our previous studies (e.g., Bekki \& Tsujimoto 2012, BT12).
Since the basic equations and the details of the models are already given in BT12,
we here briefly describe the models. One major difference in the present model is that
chemical yields of massive stars and CCSNe predicted from 
Limongi \&  Chieffi (2018, LC18) rather than
those from Tsujimoto et al. (1995) used in BT12
are newly adopted in the present study.
Accordingly, massive stars with $m \ge 25 {\rm M}_{\odot}$ cannot explode as CCSNe in
LC18, because they become stellar mass black holes through direct gravitational collapse.
Such a small upper bound on the mass of
CCSN progenitor stars is supported by various aspects 
including the observations (e.g., Smartt 2015), 
theoretical modeling of supernovae (e.g., Sukhbold et al. 2016),
and Galactic chemical evolution (Tsujimoto 2023).
Thus, 
chemical pollution by CCSNe can occur only after massive stars with $m<25 {\rm M}_{\odot}$
die away so that stellar winds of massive stars can pollute gas longer in the present models.

The present study with yields from LC18 assumes that
stellar winds from massive stars with $8 \le  m/{\rm M}_{\odot} \le 120$ 
and ejecta from CCSNe with $m <  25{\rm M}_{\odot}$ can chemically enrich
gas. Therefore, interstellar medium can be enriched by stellar winds
of massive OB and WR stars for a significantly longer timescale 
(until stars with $m =  25{\rm M}_{\odot}$ explode as CCSNe) so that
the nitrogen abundance can become rather high for a range of IMFs.
If we assume that CCSNe from stars with $8 \le  m/{\rm M}_{\odot} \le 120$
can all contribute to chemical enrichment, as in BT12, then
the timescale for GN-z11 to have a high nitrogen abundance becomes
very short ($\approx 10^6$ yr), because CCSNe can
rapidly lower the nitrogen abundance (e.g., Watanabe et al. 2023).
Gas from winds and CCSNe is recycled into interstellar medium just after
the gas ejection.  
Since the best-fit age to observational data is $\log_{10} {\rm (age/yr)}=6.57^{+0.09}_{-0.2}$
(S23), 
we stop the calculations at $T=10^7$ yr for all models except the ``dwarf'' model
M10 (later described).

The following IMF is adopted in the present study:
\begin{equation}
\Psi (m) = C_0 m^{-\alpha},
\end{equation}
where $m$ is the  mass of
each individual star and $\alpha$ is the IMF slope: $\alpha =2.35$
corresponds to the canonical Salpeter IMF  (Salpeter 1955).
The normalization factor $C_0$ is a function of $\alpha$,
the lower mass cut-off (fixed at $0.1 {\rm M}_{\odot}$),
and the upper mass cut-off ($120 {\rm M}_{\odot}$).
IMF-averaged yields for stellar winds and CCSNe are separately calculated for a given $\alpha$
using ``rotating'' models with rotational velocities of 300 km s$^{-1}$
and metallicities ($Z$) of $0.01 Z_{\odot}$ by LC18.

The total masses of gas ($M_{\rm g}$) and stars ($M_{\rm s}$) evolve with time due to
gas accretion and gas consumption by star formation. The gas infall time scale
($t_{\rm inf}$) is a parameter ranging from 1 Myr to 1 Gyr (BT12), and the
star formation rate SFR ($\psi(t)$) is assumed to be proportional
to the gas fraction ($f_{\rm g}$) as follows:
\begin{equation}
\psi(t)=C_{\rm sf}f_{\rm g}(t),
\end{equation}
where $C_{\rm sf}$ controls the timescale of star formation ($t_{\rm sf}$),
which is derived by dividing $M_{\rm g}$ by SFR at the final time step.
Although ejecta from all CCSNe is assumed to be retained within
massive dwarf  galaxies ($M_{\rm s} \approx 3 \times 10^9 {\rm M}_{\odot}$)
in BT12,
we  consider that GN-z11
with the assumed $M_{\rm s} \approx 3 \times 10^8 {\rm M}_{\odot}$
can retain only a fraction of
the ejecta.
Accordingly, the mass fraction of ejecta from  CCSNe is assumed to be
a parameter represented by $f_{\rm r}$ whereas all gas from stellar winds is
assumed to be retained.
Initial [Fe/H] in infalling gas is set to be $-2.5$ dex and
the initial [N/O] and [C/O]  are 
consistent with the observed
$\log {\rm (N/O)}$ and
$\log {\rm (C/O)}$ in low-metallicity extragalactic HII regions.

It is found that the observed gas-phase abundances and SFR
can be best reproduced by the model with $\alpha=1.55$, $t_{\rm sf}=6.6 \times 10^6$ yr
(corresponding to $C_{\rm sf}=15$),
and $t_{\rm inf}=10^6$ yr, which predicts the final SFR,
$M_{\rm s}$, and $M_{\rm g}$
at $T=10$ Myr
are $20 {\rm M}_{\odot}$ yr$^{-1}$, $3.0 \times 10^8 {\rm M}_{\odot}$,
and $1.3 \times 10^8 {\rm M}_{\odot}$, respectively.
We therefore focus on this ``fiducial'' model (M1)
with these parameter values.
We consider that the smaller $M_{\rm s}$ compared to
$\log (M_{\rm s}/{\rm M}_{\odot}) =9.1^{+0.3}_{-0.4}$ in Tacchella et al. (2023)
is an outcome of the adopted top-heavy IMF 
(and star formation history)
instead of
the canonical one that is used for estimating the observed $M_{\rm s}$.
Table 1 summarizes the parameter values for all of the ten models.


\section{Results}

Fig. 1 clearly demonstrates that
the observed location of GN-z11 on the $ 12 + \log {\rm (O/H)}$-$ \log {\rm (N/O)}$  plane 
can be reproduced well by  the models  with $\alpha=1.15$  and 1.55 (but not by the model with the
Salpeter IMF),
which suggests a top-heavy IMF
is  required for the high $\log {\rm (N/O)}$ of GN-z11.
This is essentially because the mass fractions of nitrogen-rich ejecta from
OB and WR stars in the models with top-heavy IMFs can be significant so that
the mean gas-phase nitrogen abundances can be rather high even after mixing (dilution) of 
the ejecta with infalling gas.
Fig. 1 also shows that the three models with different $\alpha$ can have
$\log {\rm (C/O)}$ higher than the observed fiducial value (Cameron et al. 2023),
which means that
the observed lower limit of $\log {\rm (C/O)}$ cannot be a strong constraint
on the model parameters.
The derived high $\log {\rm (C/O)}$ is due to the high carbon yields 
of stellar winds in LC18.
The model M1 shows $\log {\rm (N/O)} \approx -0.24$ at $T\approx 5$ Myr
(to T=10 Myr), which means that the timescale of such a high N/O  
phase is not short.
This model shows a significant increase of [Fe/H] from $-2.5$ to $-2.4$ during
its 10 Myr evolution, 
which means a larger [Fe/H] spread in the stars.

The dwarf model with $\alpha=2.35$ and $t_{\rm inf}=300$ Myr,
which represents the early star formation histories of massive dwarf galaxies
like the Large Magellanic Cloud (LMC),
does not show a high $\log {\rm (N/O)}$ even after $10^8$ yr evolution.
This result indicates that  $\log {\rm (N/O)}$ of GN-z11 is quite distinct from
those of  extragalactic HII regions due to the combination
of its very short gas infall/star formation timescales and the top-heavy 
IMF. It should be stressed here that all of the present models
show higher $\log {\rm (C/O)}$ ($>-0.4$), which means that
the present model is unable to explain lower  $\log {\rm (C/O)}$ ($<-0.4$)
observed in some of the HII regions.
It should be also noted that S23 suggested a low  $\log {\rm (C/O)}$ ($ \approx -0.5$)
for GN-z11.
The adoption of significantly lower carbon  yields for  models of  massive stars
in LC18 
would alleviate these problems of high $\log {\rm (C/O)}$ in the models.
We will discuss how this problem can be solved
in our forthcoming papers based on different models
for stellar yields from massive stars with different rotational velocities.

\begin{figure}
\psfig{file=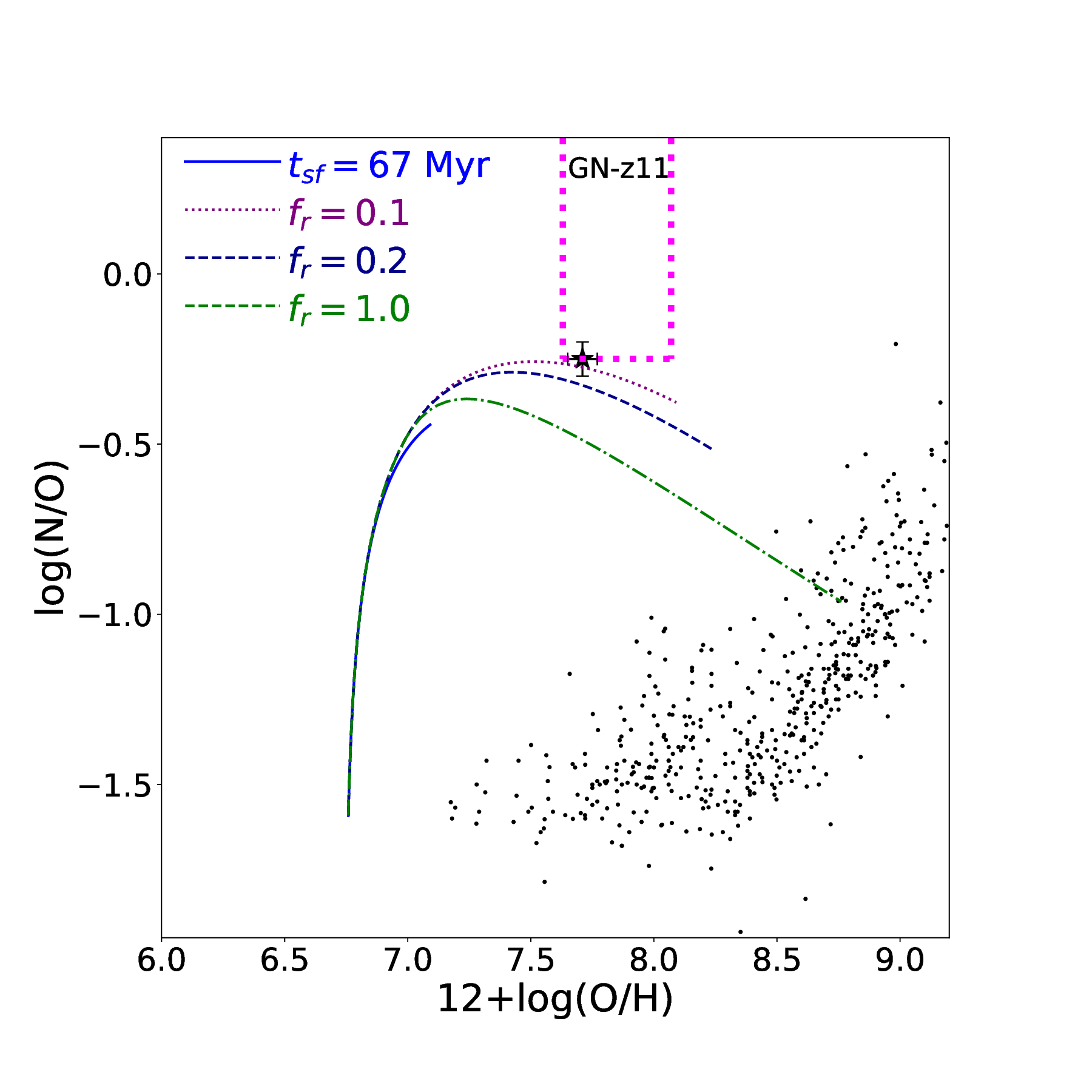,width=9.0cm}
\caption{
The same for Fig. 1 but for 
M2 with $t_{\rm sf}=6.7 \times 10^7$ yr and $f_{\rm r}=0.03$ (blue solid),
M7 with $t_{\rm sf}=6.6 \times 10^6$ yr and $f_{\rm r}=0.1$ (purple dotted),
M8 with $t_{\rm sf}=6.6 \times 10^6$ yr and $f_{\rm r}=0.2$ (dark-blue dashed), 
and M9 with 
$t_{\rm sf}=6.6 \times 10^6$ yr and $f_{\rm r}=1.0$ (green  dot-dashed) 
on the $\log {\rm 12 + (O/H)}-\log {\rm (N/O)}$ plane.
}
\label{Figure. 2}
\end{figure}

Fig. 2 shows that 
the model with longer $t_{\rm sf}$
($6.6 \times 10^7$ yr) cannot reproduce the observed high $\log {\rm (N/O)}$,
which suggests that $t_{\rm sf}$ is one of key  parameters for 
high $\log {\rm (N/O)}$ of GN-z11.
The physical reason for this $t_{\rm sf}$-dependence
is as follows.
If $t_{\rm sf}$ is very short, then a large number of massive stars
can be formed within a short time scale so that
the mass fraction of nitrogen-rich gaseous ejecta from stellar winds of the OB and WR stars 
in  all gas
can be significant.
As a result of this, gas-phase $\log {\rm (N/O)}$ can rapidly
increase before gaseous ejecta with lower nitrogen abundances
from CCSNe can start to pollute the gas.
In the models with longer $t_{\rm sf}$, on the other hand,
a larger amount of gas can be accreted and subsequently consumed very slowly by
star formation.
Consequently,  the mass fraction of nitrogen-rich ejecta from stellar winds
cannot be so high due to the dilution of the ejecta by the infalling gas.
Thus,  $\log {\rm (N/O)}$ can be rather high  only in the models with
short $t_{\rm sf}$.

As shown in Fig. 2, 
the models with  
$f_{\rm r}>0.2$  cannot have 
$\log {\rm (N/O)} \approx -0.49$,  
which corresponds to the ``conservative'' lower limit of  the observed $\log {\rm (N/O)}$
(Cameron et al. 2023): the model with $f_{\rm r} = 0.2$ can show  $\log {\rm (N/O)} \approx -0.49$.
Accordingly, we can conclude that at least 80\% of ejecta from CCSNe needs to be expelled
from GN-z11 to keep its high gas-phase nitrogen abundance. 
Fig. 3 shows that the number fraction of nitrogen-rich stars with [N/Fe]$>0.5$
depends strongly on $\alpha$ such that it is larger for more top-heavy IMFs (i.e., smaller $\alpha$).
This is firstly
 because gas-phase nitrogen abundances can become higher for more top-heavy IMFs
(due to more mass of winds),
and secondly because 
the nitrogen abundances of new stars at a given time are the same
as those of gas at that time.
The large fraction ($\approx 0.6$)  of N-rich stars 
for $\alpha=1.55$ implies that 
even the integrated spectra of GN-z11 can possibly show [N/Fe]$>0.5$.

\section{Discussion and conclusions}

The present study has demonstrated that (i)
the observed high log(N/O) of GN-z11 can be reproduced
in the model with yields from LC18 in which
OB and WR stars
and CCSNe with $m <  25{\rm M}_{\odot}$ can chemically enrich the interstellar medium
and
(ii) both top-heavy IMFs ($\alpha$ as flat as $1.5$) 
and very short $t_{\rm sf}$ ($\approx 10^7$ yr) are required to explain the observation.
Recent theoretical models for integrated galaxy-wide IMFs (``IGIMF'') predict that IMFs can be more 
top-heavy for higher SFRs (e.g., Yan et al. 2017).
Therefore, the required top-heavy IMFs are consistent with the IGIMF theory 
for GN-z11 with a rather high SFR of $\approx 20 {\rm M}_{\odot}$ yr$^{-1}$, at least
qualitatively. However, it is yet unclear why the required short $t_{\rm sf}$ is possible
in GN-z11 at $z=10.6$. 
S23 pointed out that the emission line properties of GN-z11 are
strikingly similar to those of the blue compact dwarf (BCD) galaxy  Mrk 996.
Given that BCDs can be formed from dwarf-dwarf merging (e.g., Bekki 2008; Chhatkuli et al. 2023),
high-$z$ merging of low-mass dwarfs might be responsible for the required short $t_{\rm sf}$:
it should be noted here that a candidate of such high-$z$ ($z$=10.17) merging has been recently identified by
Hsiao et al. (2023) with JWST.
The ``Haze'' observed around GN-z11 (Tacchella et al. 2023)
might be possible evidence for such a 
merger event.

\begin{figure}
\psfig{file=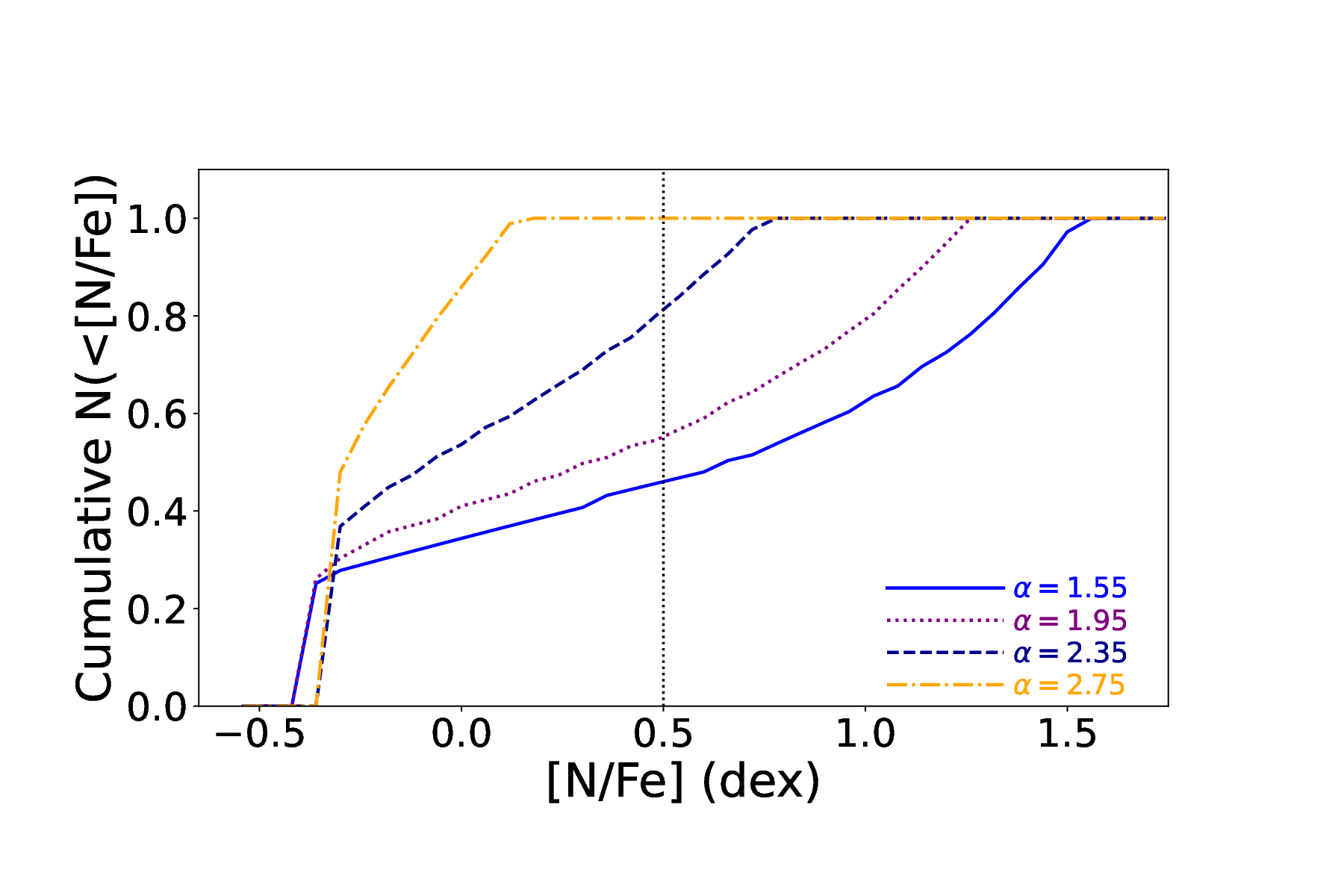,width=8.5cm}
\caption{
Cumulative distributions of [N/Fe] for stars in the four models with different IMF slopes,
$\alpha=1.55$ (M1, blue solid),
1.95 (M4, purple dotted),
2.35, (M5, dark-blue dashed),
and 2.75 (M6, orange dot-dashed).
The vertical dotted line indicates the threshold above which stars can be identified as
nitrogen-rich (Schiavon et al. 2017): M60-UCD1 has [N/Fe]=0.61 and
the Galactic N-rich stars have $0.5 \le {\rm [N/Fe]} \le 1.1$.
}
\label{Figure. 3}
\end{figure}

If the IMF of GNz11 is indeed top-heavy, 
$M_{\rm s}$ of GN-z11 estimated in previous studies based on the canonical IMF 
(Tacchella et al. 2023)
could  be an overestimation.
The much larger numbers of CCSNe and prompt SNIa for top-heavy IMFs imply
very efficient energetic feedback effects of supernovae (SNe) that can  truncate star formation
very rapidly. It is therefore likely that GN-z11 can evolve into a poststarburst system
after its SNe expel the remaining gas almost completely, if no further gas supply/accretion
is possible.
Recently,  Strait et al. (2023) have discovered a very compact ($R_{\rm h} \approx 30$pc)
post-starburst galaxy at $z=5.2$ (MACS0417-z5PSB):
there could be an evolutionary link between very compact post-starburst galaxies
like MACS0417-z5PSB and very compact star-forming ones with high nitrogen abundances
like GN-z11 in the high-$z$ universe.

If GN-z11 with $M_{\rm s}=3 \times 10^8 {\rm M}_{\odot}$ 
($10^9 {\rm M}_{\odot}$)  at z=10.6
is formed with $\alpha=1.55$ and if it stops its ongoing star formation due to feedback
effects of numerous CCSNe and
consequently evolves passively until now,
then its present-day total mass of low-mass stars ($m<0.8 {\rm M}_{\odot}$) is
only $2.0 \times 10^7 {\rm M}_{\odot}$
($6.6 \times 10^7 {\rm M}_{\odot}$).
These masses are more consistent with those of UCDs and stellar nuclei in massive 
dwarfs (e.g., C\^ote et al. 2006) than those of GCs ($\approx 2 \times 10^5 {\rm M}_{\odot}$).
If UCDs 
with $M_{\rm s}$ at  $z=0$ larger than $2 \times 10^7 {\rm M}_{\odot}$
like M60-UCD1 are formed at $z>10$,
they should be significantly brighter than GN-z11 in the rest-frame UV wavelength: 
they will be able to be detected in
JWST observations. 
Given that the number fractions
of nucleated dwarfs are higher in dense cluster environments (e.g., C\^ote et al. 2006),
the discovery of such nitrogen-rich compact objects  like GN-z11
can indicate the central regions of proto clusters of galaxies
(see Tacchella et al. 2023 for such possible evidence).

If a significant number of high-$z$ objects with high $\log {\rm (N/O)}$ and 
$M_{\rm s}=[10^6-10^7] {\rm M}_{\odot}$ are discovered,
their physical properties can be discussed in the context of
(i) self-enrichment of forming GCs with multiple stellar populations
through stellar winds of massive stars
(e.g., Prantzos \& Charbonnel 2006) and (ii) the Galaxy formation
(e.g., Belokurov \&  Kravtsov 2023).
It should be stressed here that stellar populations of GN-z11 can
have large [Fe/H] spreads ($>0.05$ dex) due to the chemical enrichment
by low-mass ($m<25 {\rm M}_{\odot}$) CCSNe: GN-z11 could differ from
normal GCs with small [Fe/H] spreads ($<0.05$ dex: Carretta et al. 2019).
GN-z11 also might  have (i) no 
O-Na and Mg-Al anti-correlations observed in  GCs (e.g., Carretta et al. 2019)
and (ii)  smaller [Na/O] in its integrated spectra (compared to GCs)
if it was chemically enriched both by OB and WR stars and by low-mass CCSNe.

The Galaxy is observed to have a significant fraction of ``N-rich'' stars with [N/Fe]$>0.5$
(e.g., Schiavon et al. 2017),  the physical origin of which is yet to be fully understood in
theoretical models of galaxy formation (e.g., Bekki 2019). 
Given that even the integrated spectra of elliptical galaxies show
moderately high [N/Fe] (up to $ \approx 0.2$;
Schiavon 2007), they should contain a significant fraction of N-rich stars. 
Large fractions of N-rich stars derived in the present models with top-heavy IMFs
suggest that if some of the low-mass building blocks of the Galaxy and elliptical galaxies
are like GN-z11, the galaxies should be able to contain N-rich stars 
after the tidal destruction of the building blocks.

The present model predicts a large number ($>10^6$) 
of stellar mass BHs within the central $\approx 60$pc of GZ-z11,
where there should be a plenty of cold gas to fuel ongoing star formation:
the possible presence of active galactic nucleus 
(Maiolino et al. 2023) might be 
physically related to such a  BH cluster.
The retention probability of stellar mass BHs in massive stellar systems
with $M_{\rm s}>10^7 {\rm M}_{\odot}$ is almost 100\%
(Je{\v{r}}{\'a}bkov{\'a} et al. 2017).
Therefore,
dynamical evolution of such a dense cluster of BHs within a 
massive gas-rich environment is a new
dynamical problem, 
because both two-body BH-BH interaction and hydrodynamical interaction
between gas and BHs are quite important for the evolution of the self-gravitating system.
Our future studies on the dynamical evolution of BH clusters in gas-rich environments 
will be able to address whether or not there is a physical link between dense clusters of BHs
in forming compact galaxies  at high $z$
and MBHs in the present-day UCDs and stellar nuclei.

\section{DATA AVAILABILITY}
The data used in this paper (outputs from  one-zone models)
will be shared on reasonable request
to the corresponding author.

\section{Acknowledgment}
We are   grateful to the referee  for  constructive and
useful comments that improved this paper.
T. T. acknowledges the support by JSPS KAKENHI Grant Nos. 18H01258, 19H05811, and 23H00132.

\end{document}